\documentclass[journal]{IEEEtran}
\usepackage{times}
\usepackage{epsfig}
\usepackage{graphicx}
\usepackage{amsmath}
\usepackage{amssymb}
\usepackage{array}
\usepackage{multirow}
\usepackage{subfigure}

\newcommand{\etal}{\textit{et al}.}
\newcommand{\ie}{\textit{i}.\textit{e}.}

\usepackage[pagebackref=true,breaklinks=true,letterpaper=true,colorlinks,citecolor=red,linkcolor=red,bookmarks=false]{hyperref}

\begin{document}
%
\title{PROGRESSIVE RESIDUAL LEARNING FOR SINGLE IMAGE DEHAZING }

%
%
%

\author{Yudong~Liang,
        Bin Wang,
        Jiaying Liu,~\IEEEmembership{Senior Member,~IEEE,}
        Deyu Li,
        Yuhua Qian,       
        and~Wenqi Ren,~\IEEEmembership{Member ,~IEEE,}
\thanks{Yudong Liang, Bin Wang, Deyu Li, Yuhua Qian are School of Computer and Information Technology, Shanxi University, Shanxi, China.}
\thanks{Jiaying Liu is with the Institute of Computer Science and Technology, Peking University, Beijing, China.}
\thanks{Wenqi Ren is with State Key Laboratory of Information Security, Institute of Information Engineering, Chinese Academy of Sciences.}}

\markboth{Journal of \LaTeX\ Class Files,~Vol.~14, No.~8, August~2015}%
{Shell \MakeLowercase{\textit{et al.}}: Bare Demo of IEEEtran.cls for IEEE Journals}
%



\maketitle

\begin{abstract}
 The recent physical model-free dehazing methods have achieved state-of-the-art performances. However, without the guidance of physical models, the performances degrade rapidly when applied to real scenarios due to the unavailable or insufficient data problems. On the other hand, the physical model-based methods have better interpretability but suffer from multi-objective optimizations of parameters, which may lead to sub-optimal dehazing results. In this paper, a progressive residual learning strategy has been proposed to combine the physical model-free dehazing process with reformulated scattering model-based dehazing operations, which enjoys the merits of dehazing methods in both categories. Specifically, the global atmosphere light and transmission maps are interactively optimized with the aid of accurate residual information and preliminary dehazed restorations from the initial physical model-free dehazing process. The proposed method performs favorably against the state-of-the-art methods on public dehazing benchmarks with better model interpretability and adaptivity for complex hazy data.
\end{abstract}

\section{Introduction}
\label{sec:intro}
Image dehazing is essential for improving the visibility of the hazy images as well as facilitating the following vision-based applications.
The existed image dehazing approaches can be roughly classified into two categories: physical model-based and model-free image dehazing methods.

The physical model-based dehazing methods follow the law of light propagation to restore the dehazed images by estimating the parameters of the models. 
The most widely applied physicals model in the literature is the atmospheric scattering model \cite{he2010single,ren2016single,zhang2018densely}, which describes the formulation of the hazy image $I(x)$ as: 
\begin{equation}\label{eq:scatter}
    \begin{aligned}
    I(x)=J(x)t(x)+A(1-t(x)),
    \end{aligned}
 \end{equation}
The scene radiance or the hazy-free image $J(x)$ decays in the medium, while the portion of the light that is not scattered and reaches the camera is decided by the transmission map $t(x)$. The airlight $A(1-t(x))$ is controlled by the global atmospheric light $A$ which is supposed to be homogeneous. Scattering model-based dehazing methods estimate global atmospheric light and transmission maps and then inverse the physical model to recover the hazy-free image $J(x)$.

The estimation of the transmission map is the main focus of the physical model-based dehazing methods. Following the atmospheric scattering model, He \etal~\cite{he2010single} proposed a dark channel prior to estimate the transmission map.
Ren \etal~\cite{ren2016single} predicted the transmission map via a deep network with the aid of multi-scale information. Zhang and Patel~\cite{zhang2018densely} embedded the atmospheric scattering model into a Densely Connected Pyramid Dehazing Network, termed DCPDN, to estimate transmission map, global atmospheric light and hazy-free images.
\begin{figure*}[th]
     \centering
     \footnotesize
    \includegraphics[width=0.8\textwidth]{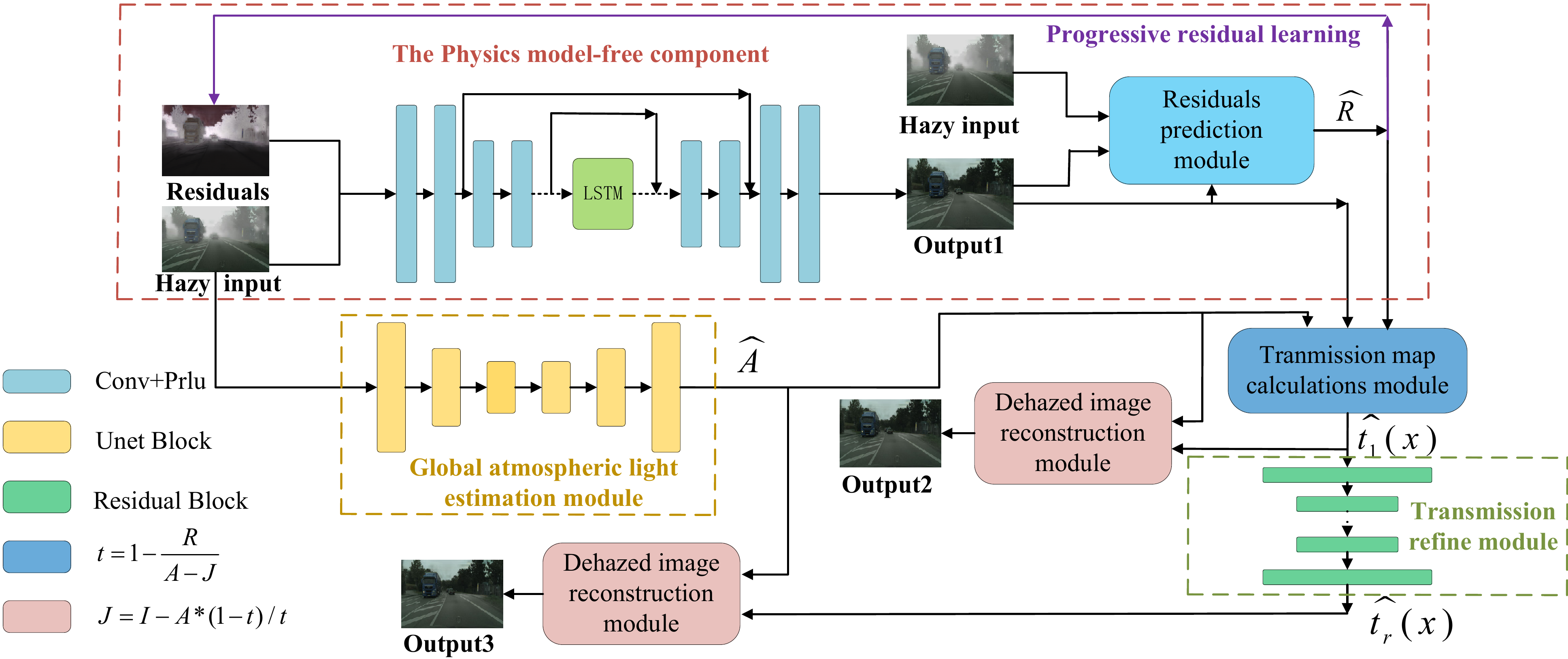}
    \caption{The framework of the proposed approach.}
  \label{fig:arch}
  \end{figure*}
 Although the atmospheric scattering model applies well in many situations and has better model interpretability, small deviations from the estimation of either transmission map or global atmospheric light can severely damage the dehazed images even if another parameter is accurate. Multi-objective optimizations and non-interactive predicting of two critical parameters (transmission map and atmospheric light) may lead to sub-optimal learning of final dehazed results. 

 On the other side, model-free dehazing methods~\cite{ren2018gated,qu2019enhanced,dong2020multi} are directly optimized to restore the hazy-free images without considering the light scattering rules. Without multi-step or multi-objective optimizations for predicting transmission maps, global atmosphere light, and hazy-free images, the recent physical model-free dehazing methods \cite{qu2019enhanced,dong2020multi} have achieved state-of-the-art performances on synthetic datasets.
 However, without the guidance of the physical models, the performances degrade rapidly when the methods are applied to real scenarios due to the unavailable or insufficient data problems.

It is intuitive to make efforts to take advantage of methods in both categories.
Derived from Eq.~\ref{eq:scatter}, the calculations of transmission map, global atmospheric light and  residuals information are interactive in dehazing process by the physical model method, which can be explained as follow:
 \begin{equation}\label{eq:Trans}
    t(x)= \frac{I(x)-A}{J(x)-A}=1-\frac{J(x)-I(x)}{J(x)-A}=1-\frac{R}{J(x)-A}.
 \end{equation}
where residuals refer to the deviations between the haze-free image and the hazy input. The residuals information can be easily obtained in a physical model-free dehazing method.
Accurate residuals can clearly benefit the estimation of the transmission map and the following dehazing process. Physical model-free and model-based dehazing processes are thus elegantly connected by the residuals learning.

In this paper, we attempt to combine the advantages of both model-based and model-free methods to improve the performance, interpretability, robustness, and adaptivity of the proposed model. A progressive residual learning strategy has been utilized to cascade the physical model-free dehazing process with reformulated scattering model-based dehazing operations.
To deal with the information memorizing and forgetting and gradient vanishing problems, LSTM is utilized during the progressive residual learning process. Residuals between dehazed images and hazy inputs are iteratively refined to facilitate the following estimating of transmission map. Better dehazed results would benefit the residuals information predictions and vice versa.

In the reformulated scattering model based dehazing process, global atmosphere light and transmission maps are interactively optimized as transmission map can be better estimated with the aid of the residual learning.
With the preliminary 
residuals predictions, more accurate and robust estimations of global atmospheric light and transmission maps can be obtained from the reformulated scattering model than the conventional physical model-based methods. The experiments demonstrate the proposed progressive residual learning strategy not only brings better accuracy and adaptivity for the dehazing task,
but also performs well on real hazy data.

\section{Related work}
Learning residuals has largely improved the performances in many low-level vision tasks. Kim \etal \cite{kim2016accurate} proposed to learn the residuals between high-resolution and low-resolution images for image super-resolution problem, which largely accelerated the training speed and performances of deep networks. Zhang \etal \cite{zhang2017beyond} proved high effectiveness in general image denoising tasks with residual learning.
Inspired by the existed successes of learning residuals, the progressive residual learning strategy is applied to combine the physical model-free and reformulated scattering model-based dehazing component into a deep end-to-end model.

\section{Proposed method}
Our proposed model is an end-to-end dehazing model which cascades model-free dehazing component and reformulated scattering model-based component with a progressive residual learning strategy as Fig. \ref{fig:arch}.  The residuals refer to the estimated deviations between haze-free images and hazy inputs.

\subsection{The physical model-free dehazing component}
The physical model-free dehazing component consists of a recursive dehazing and residual learning module as Fig.~\ref{fig:arch}.
The component is fed with hazy input $I$ and previously estimated residuals $\hat{R}_{k-1}$ to estimate current dehazed output $\hat{J}_{k}$ and current residuals $\hat{R}_{k}$ iteratively via a Long short-term memory (LSTM) network. The current estimated residuals $\hat{R}_{k}$ are calculated as a function of deviations between the current dehazed output $\hat{J}_{k}$ and hazy input $I$ as Eq.~\ref{eq:predictR}:
 \begin{equation}\label{eq:predictR}
    \hat{R}_k=f_{R}(I-\hat{J}_{k}),
    \end{equation}
where the function $f_{R}$ is learned by a deep module. Dehazed output $\hat{J}_{k}$ is predicted by a lightweight deep convolutional network directly from hazy input and previously estimated residuals. All the values of the initial estimated residuals map $\hat{R}_{0}$ are set to zero. Optimizing two variable (\ie dehazed output and residuals) alternately is easy to become unstable or get stuck into local minimal. With the LSTM deep networks, gradients problem is alleviated and the dehazed output and estimated residuals are iteratively refined. Better dehazed output would facilitate more accurate residuals estimations and vice versa.

\subsection{The scattering model-based dehazing component}
Then the reformulated scattering model-based dehazing network is cascaded with the previous physical model-free dehazing component. The hazy input, dehazed output and estimated residuals by the previous part are imported to the scattering model-based dehazing network to progressively refine the estimation of transmission map.

A transmission prediction deep module is constructed to estimate the transmission map $\hat{t(x)}$ from the estimated residuals $\hat{R}$, estimated global atmospheric light $\hat{A}$ and previous dehazed image $\hat{J}$ as Eq. \ref{eq:predictT}.
 \begin{equation}\label{eq:predictT}
    \hat{t(x)}=1-\frac{\hat{R}}{\hat{J(x)}-\hat{A}}.
 \end{equation}
Then a transmission refinement deep module which consists of residual blocks is built to further improve the estimation of the transmission map denoted as $\hat{t}_{refine}$.

The global atmospheric light estimation module utilizes a conventional encoder-decoder deep module.
The existed scattering model-based methods utilize the groundtruth global atmospheric light, transmission map and dehazed image to supervise the learning of the physical model. Although dehazed images are synthesized according to the global atmospheric light and transmission map, multi-objective optimizations and non-interactive predicting of two critical parameters as well as dehazing images may impact the performances of the model. Small deviations of either global atmospheric light or transmission map may severely damage the performances.
To improve the robustness of the scattering model based methods, the global atmospheric light is not separately optimized and no groundtruth global atmospheric light is provided. Instead, the global atmospheric light is interactively optimized and supervised from the transmission map calculations as the transmission map is calculated according to the global atmospheric light and residuals information.
The global atmospheric light is explicitly supervised from the estimations of the transmission map and dehazed images which ensures the learned atmospheric light and transmission map can cooperate with each other to restore the desired dehaze images.
\begin{figure*}[ht]
\setlength{\tabcolsep}{1pt}
\centering
\footnotesize
\resizebox{1.0\linewidth}{!}
{
\centering
\begin{tabular}{cccccc}
\includegraphics[width=0.16\linewidth]{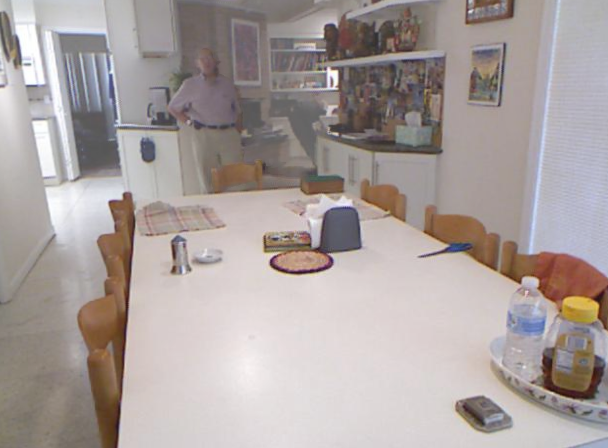}&
\includegraphics[width=0.16\linewidth]{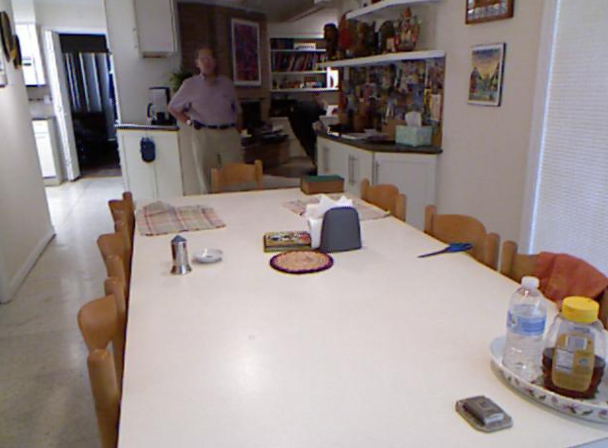}&
\includegraphics[width=0.16\linewidth]{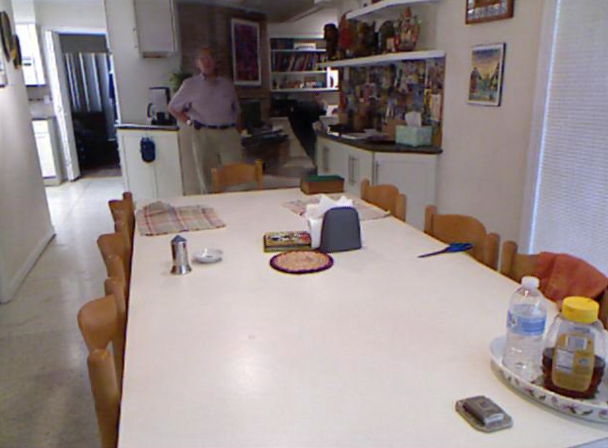}&
\includegraphics[width=0.16\linewidth]{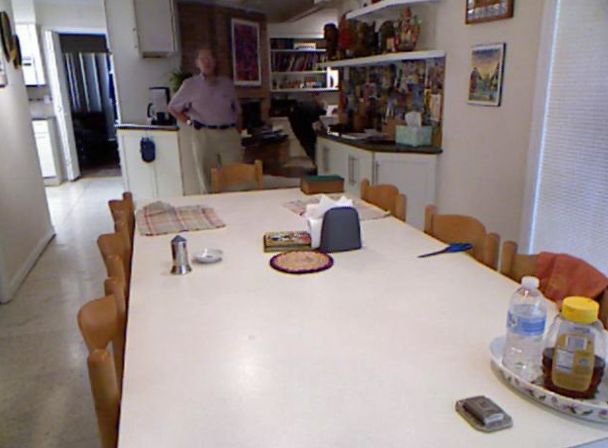}&
\includegraphics[width=0.16\linewidth]{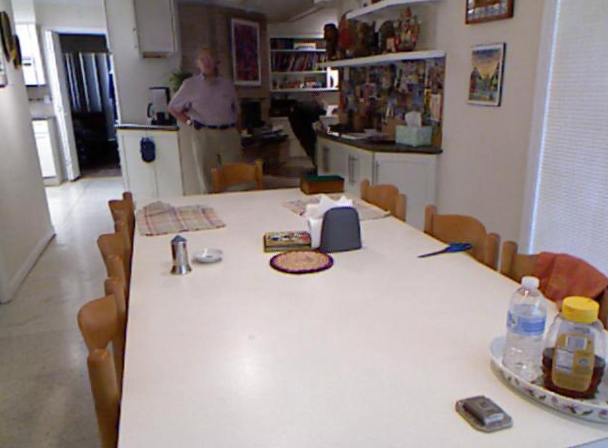}&
\includegraphics[width=0.16\linewidth]{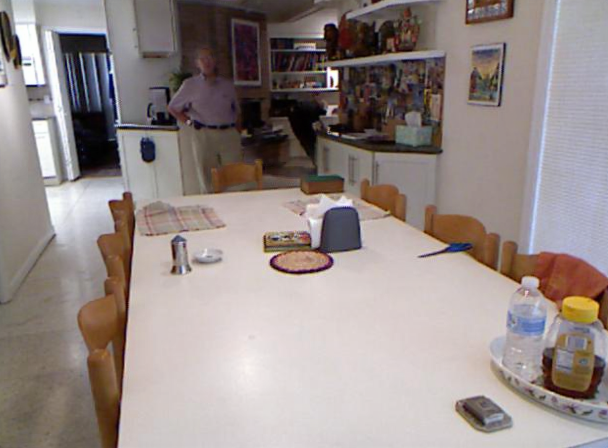}\\

 Input & GT &GCANet~\cite{chen2018gated}  &GridDehazeNet~\cite{liuICCV2019GridDehazeNet}&  MSBDN~\cite{dong2020multi} & \textbf{PRdehaze}(ours) \\
\end{tabular}
}
\caption{Dehazing comparisons for the indoor images.}
  \label{fig:indoor}
\end{figure*}
\vspace{-0.5cm}
\subsection{Loss function}

To train our proposed progressive residual learning framework, the overall loss function $L$ is a summation of the loss function $L_{u}$ for the physical model-free dehazing component and the loss function $L_{v}$ for reformulated scattering model-based dehazing component.

Conventional pixel-wise reconstruction loss $L_{p}$, perceptual loss $L_{f}$ and adversarial loss $L_{adv}$ are applied for the physical model-free dehazing part as follow:
\begin{equation}\label{eq:lossu}
\begin{aligned}
    loss_{u}=  \sum_{k=1}^{K} (& \alpha_1 L_{p}(J,\hat{J}_k)+\alpha_2 L_{f}(J,\hat{J}_k)+\alpha_3 L_{adv}(J,\hat{J}_k)\\
     &+\alpha_4  L_{p}(I-J,\hat{R}_k)+ \alpha_5 L_{p}(I-\hat{J}_k,\hat{R}_k))
       \end{aligned}
 \end{equation}
 where $\alpha_i$ is the corresponding weight and K is the iteration number of residual learning for the physical model-free dehazing component. The pixel-wise reconstruction loss $L_{p}$ can be $L_{1}$ or $L_{2}$ loss. The prediction of the residuals is optimized to approximate the groundtruth residuals information (the 4th term in Eq.~\ref{eq:lossu}) and the residuals information should revise the dehazing image prediction $\hat{J}_k$ (the 5th term in Eq.~\ref{eq:lossu}).
  The perceptual loss $L_{f}$ utilizes the conventional VGG feature loss while adversarial training plays the min-max game:
 \begin{equation}\label{eq:lossadv}
     \begin{aligned}
   L_{adv}(J,\hat{J}_k))=&\min_{G_{k}}\max_{D}\mathbb{E}_{I\sim_{Pdata(I)}}\log(1-D(G_{k}(I)))\\
   &+\mathbb{E}_{J\sim _{Pdata(J)}}\log(D(J)),\\
  where\quad \hat{J}_k=& G_{k}(I).
  \end{aligned}
 \end{equation}
For the reformulated scattering model-based dehazing componnet, guidance for the transmission map and the dehazing image forms the loss function $loss_{v}$:
 \begin{equation}\label{eq:lossv}
    loss_{v}= L_{p}(\hat{t},t) + L_{p}(\hat{t}_{refine},t)+L_{p}(\hat{J},J) +L_{p}(\hat{J}_{refine},J).
 \end{equation}
 $\hat{t}$ is the preliminary transmission map estimation from Eq.~\ref{eq:predictT} and $\hat{t}_{refine}$ is a further refined transmission estimation by the following transmission refinement module.
 The global atmospheric light is interactively optimized via the calculations of the transmission map and the dehazed image.
The pixel-wise reconstruction loss functions in Eq.~\ref{eq:lossv} are calculated to penalize the deviations between estimated and groundtruth transmission map as well as deviations between estimated and groundtruth dehaze images.
$\hat{J}$ or $\hat{J}_{refine}$ is the corresponding restoration from the estimation of global atmospheric light $\hat{A}$ and transmission map $\hat{t}$ or $\hat{t}_{refine}$ respectively. 

\vspace{-0.5cm}
\subsection{Details of the architecture}
The proposed model applies convolution and PRelu layers without batch normalization operations. In the physical model-free dehazing component network, both the recursive dehazing and residual learning module apply a encoder-decoder network with skip-connections between encoder and decoder. Lstm unit is utilized in the encoder-decoder network to alleviate the gradient problem during training as \cite{li2020single}. In the reformulated physical model-based dehazing network, the global atmospheric light estimation network utilizes a U-net structure while transmission refinement network uses several residual blocks to refine the transmission map estimations.
%
%
\begin{table*}[hbt]
\begin{center}
 \caption{Quantitative PSNR / SSIM comparisons on Synthetic Objective Testing Set(SOTS) of RESIDE dataset.}
  \label{tab:SOTS}
    \scriptsize
  \begin{tabular}{ccccccccc}
    \hline
                 &\textbf{DCP}~\cite{he2010single}&\textbf{AOD-Net}~\cite{li2017aod}& \textbf{MSCNN}~\cite{ren2016single}&\textbf{GCANet}~\cite{chen2018gated}&  \textbf{GFN}~\cite{ren2018gated}&  \textbf{GridDehazeNet}~\cite{liuICCV2019GridDehazeNet} 
                 &\textbf{MSBDN}~\cite{dong2020multi} &  \textbf{PRdehaze}\\
    \hline
     PSNR / SSIM & 16.95/0.792 &  19.06/0.0.850 & 17.57 / 0.810&30.00/0.959 &24.11 / 0.899&32.08 / \textcolor[rgb]{1.00,0.00,0.00}{0.983}& 32.77/0.980 & \textcolor[rgb]{1.00,0.00,0.00}{ 33.06/0.983}\\
    \hline
  \end{tabular}
  \end{center}
\end{table*}
\begin{table*}[hbt]
\begin{center}
 \caption{Quantitative PSNR / SSIM comparisons on Foggy Cityscapes dataset.}
  \label{tab:City}
    \scriptsize
  \begin{tabular}{cccccccc}
    \hline
                 &\textbf{AOD-Net}~\cite{li2017aod}& \textbf{MSCNN}~\cite{ren2016single}& DehazeNet~\cite{cai2016dehazenet}& \textbf{GridDehazeNet}~\cite{liuICCV2019GridDehazeNet}& \textbf{SSMDN}~\cite{song2018deep}& \textbf{BidNet}~\cite{pang2020bidnet}& PRdehaze\\
    \hline
     PSNR / SSIM &  15.45 / 0.632&18.99 / 0.860 & 14.97 / 0.487& 23.72 / 0.923 & 22.38 / 0.912 &25.57 /0.944 & \textbf{28.20 / 0.971}\\
    \hline
  \end{tabular}
  \end{center}
\end{table*}
%
\begin{figure*}[hbt]
\setlength{\tabcolsep}{1pt}
\centering
\footnotesize
\resizebox{1.0\linewidth}{!}
{
\centering
\begin{tabular}{ccccc}
\includegraphics[width=0.16\linewidth]{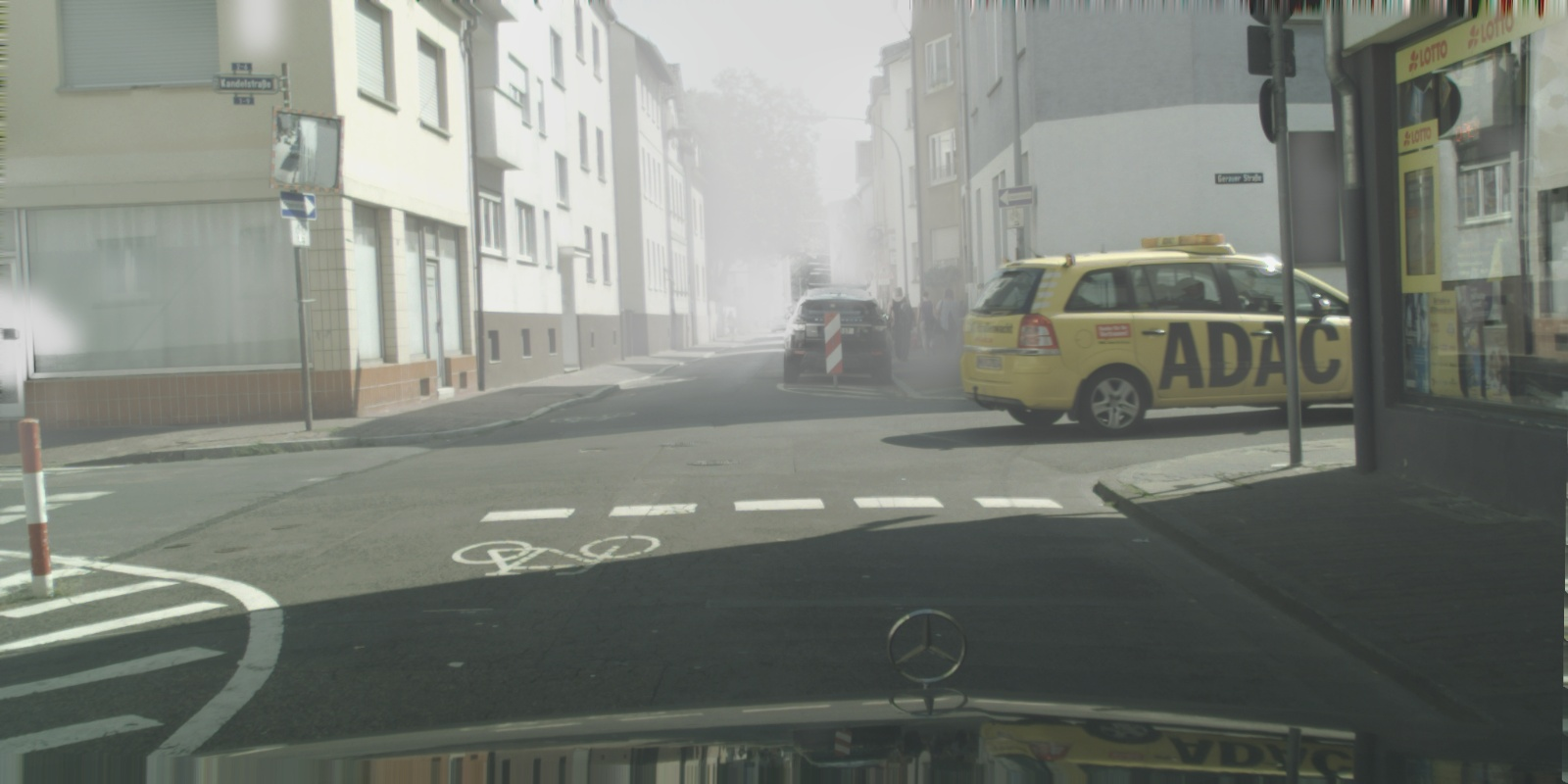}&
\includegraphics[width=0.16\linewidth]{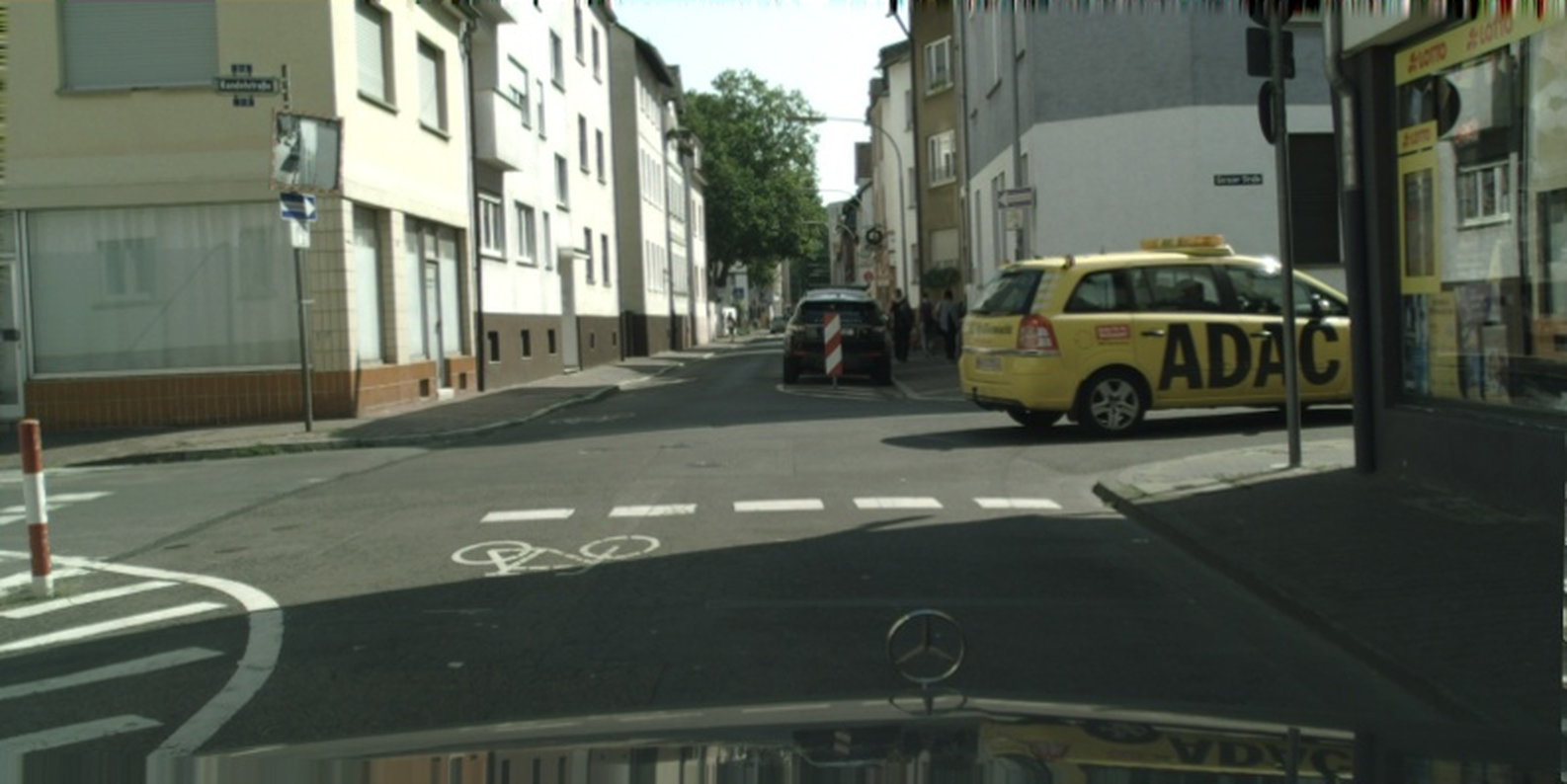}&
\includegraphics[width=0.16\linewidth]{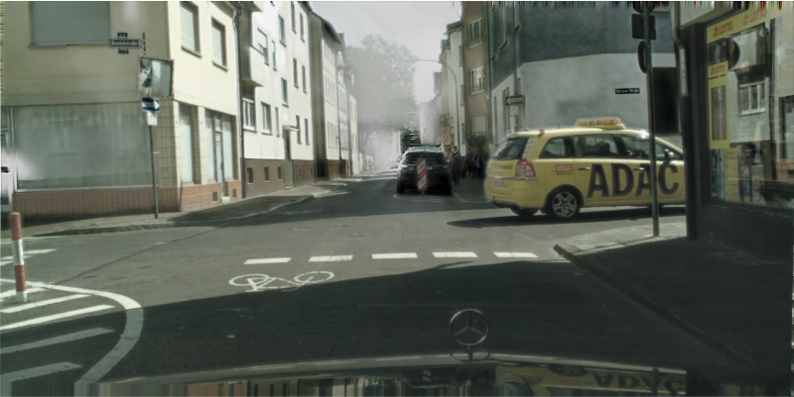}&
\includegraphics[width=0.16\linewidth]{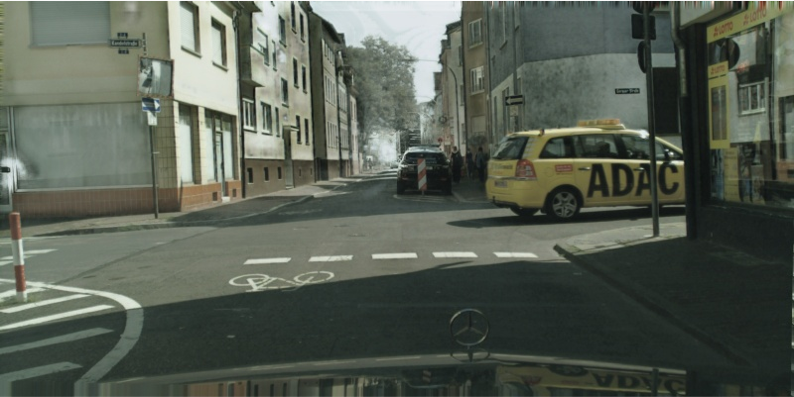}&
\includegraphics[width=0.16\linewidth]{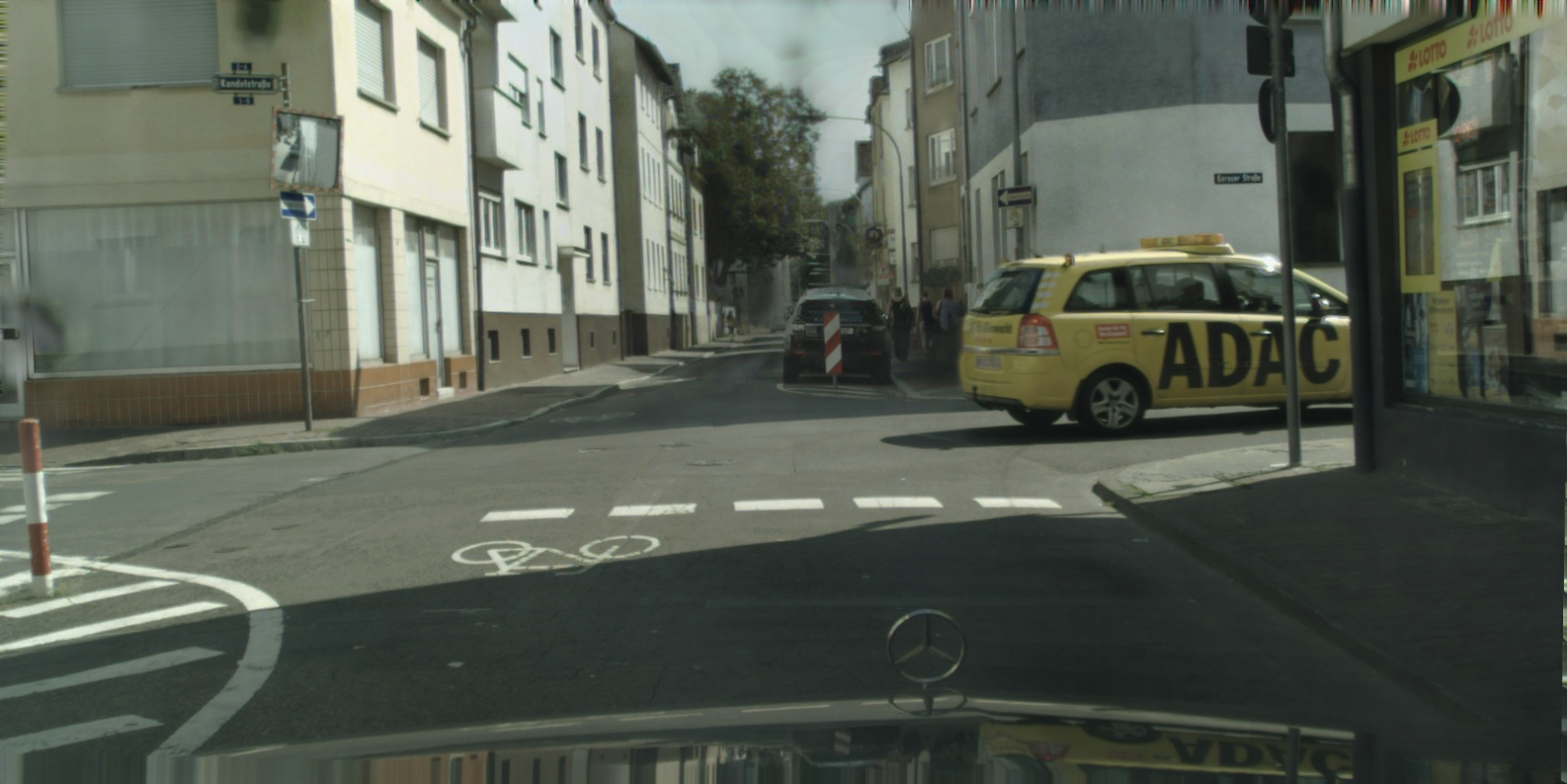}\\
\includegraphics[width=0.16\linewidth]{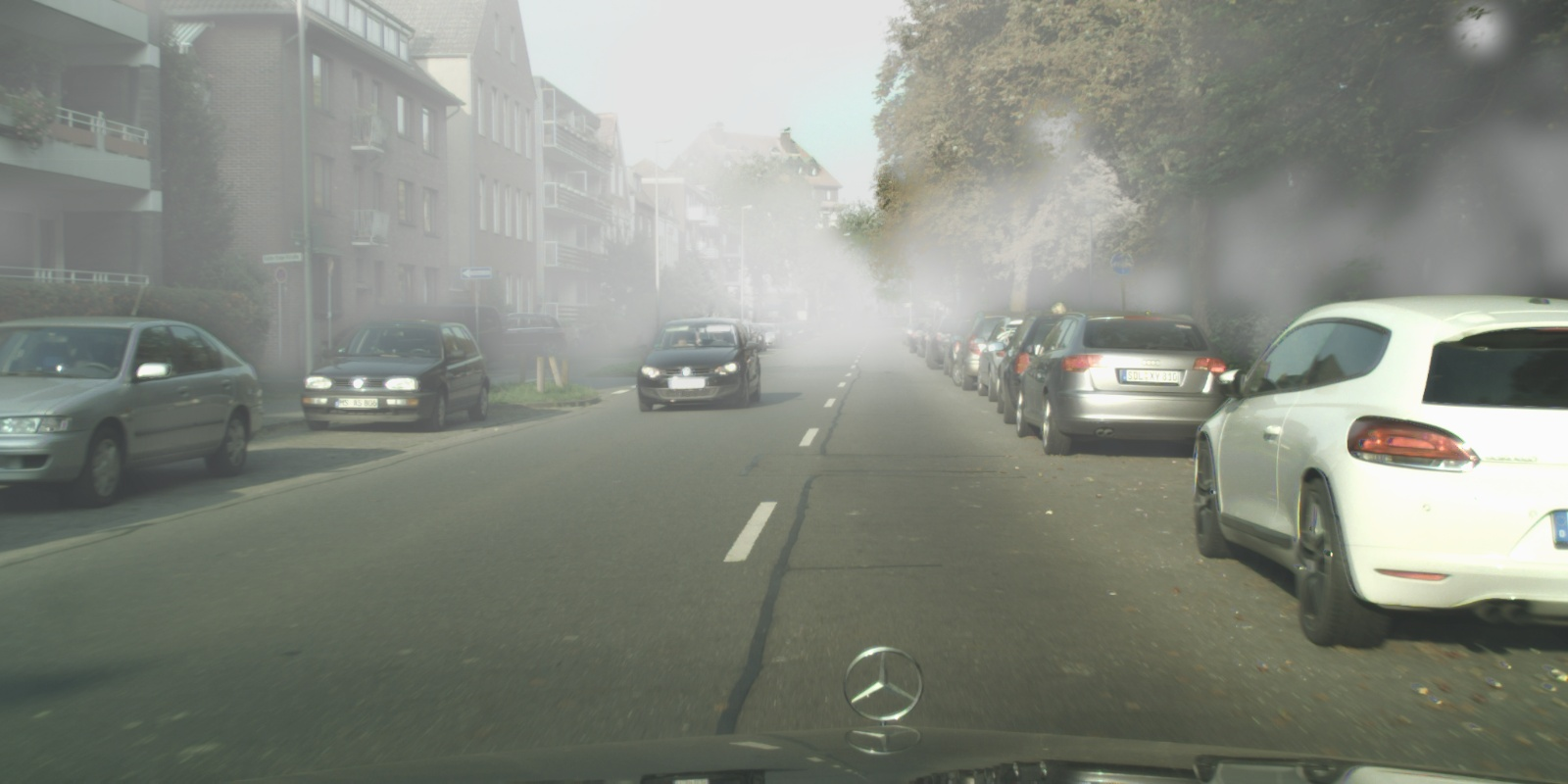}&
\includegraphics[width=0.16\linewidth]{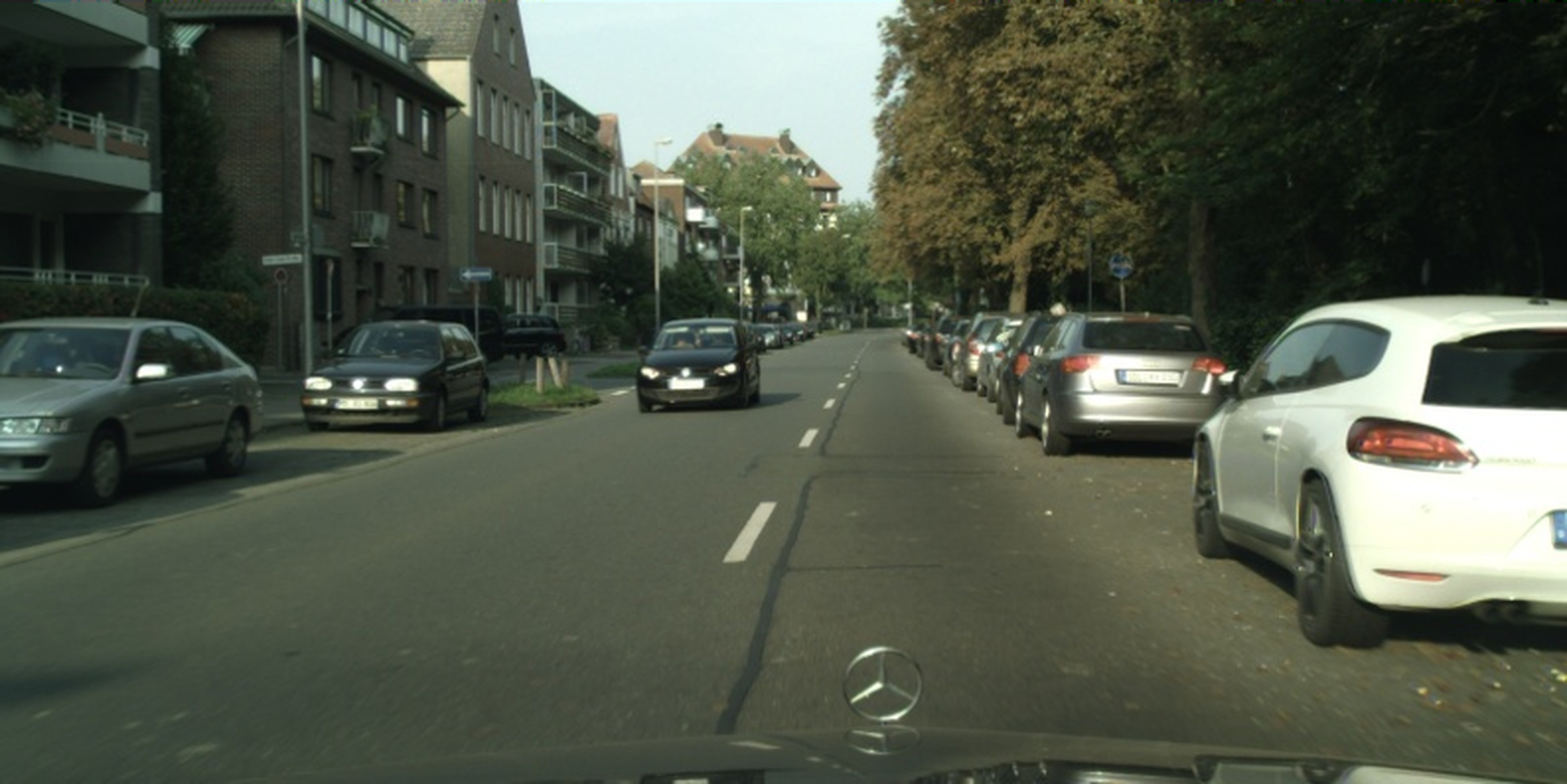}&
\includegraphics[width=0.16\linewidth]{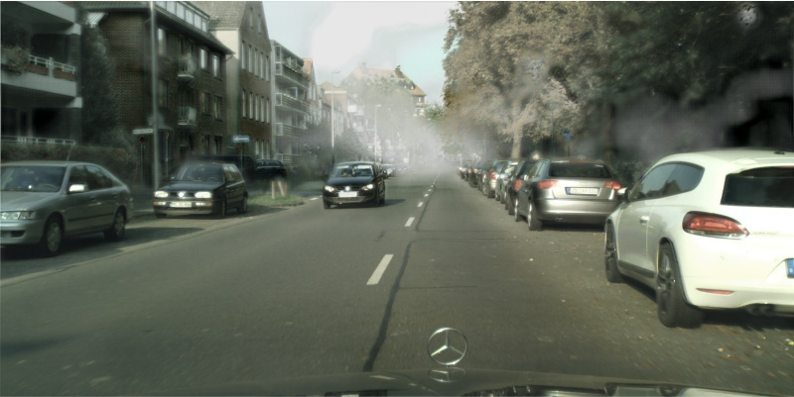}&
\includegraphics[width=0.16\linewidth]{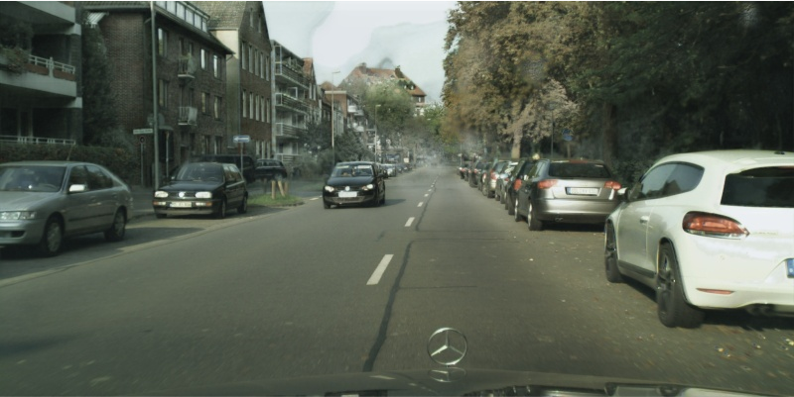}&
\includegraphics[width=0.16\linewidth]{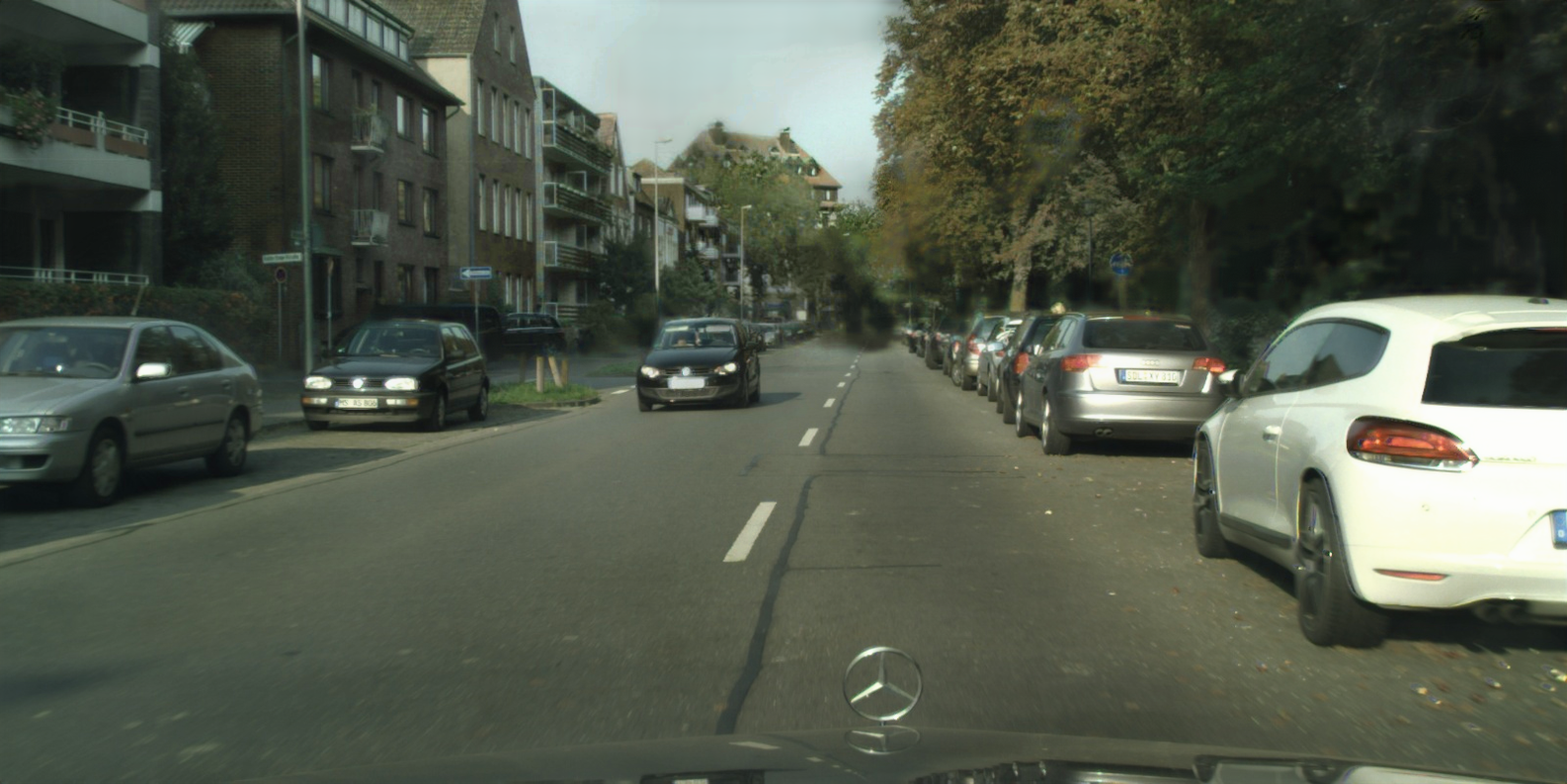}\\
 Input & GT &SSMDN~\cite{song2018deep}& BidNet~\cite{pang2020bidnet} & \textbf{PRdehaze}(ours) \\
\end{tabular}
}
\caption{Dehazing comparisons for the outdoor images.}
  \label{fig:outdoor}
\end{figure*}
\section{Experiments}
In this section, the proposed approach is compared with several state-of-the-art dehazing methods on both synthetic datasets and real-world images. For indoor images, the compared dehazing methods includ \textbf{DCP}~\cite{he2010single}, \textbf{AOD-Net}~\cite{li2017aod},  \textbf{MSCNN}~\cite{ren2016single},\textbf{GCANet}~\cite{chen2018gated},  \textbf{GFN}~\cite{ren2018gated},  \textbf{GridDehazeNet}~\cite{liuICCV2019GridDehazeNet}, 
\textbf{MSBDN}~\cite{dong2020multi}. Some state-of-the-art methods for outdoor images are utilized for the comparisons on outdoor dataset, such as DehazeNet~\cite{cai2016dehazenet}, SSMDN~\cite{song2018deep}, BidNet~\cite{pang2020bidnet}. All the compared results are either tested by the authors' released codes or otained from the reported results in their paper\footnote{Some visual results are cropped from the paper as no models are released.}. As revealed by the quantitative and visual comparisons, our progressive residual learning strategy (\textbf{PRdehaze}) performs favorably against the state-of-the-art methods.

Our model is implemented with pyTorch toolbox and the Adam optimization is utilized during the training with a RTX 2080Ti GPU.

Analogous to most of existing deep learning-based dehazing methods~\cite{cai2016dehazenet,ren2016single,li2017aod,zhang2018densely}, dehazing datasets are synthesized for the training and testing process. Further, comparisons are also performed on the real hazy image dataset~\cite{ancuti2018haze} as \cite{dong2020multi} to demonstrate the effectiveness for real applications.

\vspace{-0.6cm}
\subsection{Indoor Dataset}
The RESIDE dataset \cite{li2018benchmarking} is a large-scale benchmark and the standard version of RESIDE is applied in the experiments which contains indoor Training Set(ITS), Synthetic Objective Testing Set(SOTS), Hybrid Subjective Testing Set (HSTS). For indoor scenarios, ITS with 13990 pictures and SOTS with 500 images are utilized for training and testing respectively. The quantitative PSNR / SSIM comparisons are given as Table \ref{tab:SOTS}. The visual comparisons are represented in Fig. \ref{fig:indoor}. 
According to the Table~\ref{tab:SOTS}, our method has achieved the best performances for both PSNR or SSIM index.

\vspace{-0.3cm}
\subsection{Outdoor Dataset}
Foggy Cityscapes dataset~\cite{Sakaridis2018Semantic} contains 5000 images of urban street scenes, which splits into training, validation and testing dataset. Constant attenuation coefficients are applied to the haze-free images to determine the fog density and the visibility range. The compared methods are either finetuned or trained on the Cityscapes dataset as we did. The quantitative PSNR / SSIM comparisons are given as Table \ref{tab:City}.

Shown in the Table \ref{tab:City}, the proposed model obtains the best performances and outperforms the other methods with a large margin. In Fig.\ref{fig:outdoor}, the colors of dehazed images by our method are more natural, such as the trees or buildings.  \textbf{BidNet}~\cite{pang2020bidnet} incorrectly make the white wall of the building dark, while some areas of the dehazed images by \textbf{SSMDN}~\cite{song2018deep} still exist some haze. As the outdoor dataset is generally considered more difficult to be overfitted by the large model, our progressive residual learning improves the adaptability for the outdoor foggy images.

\begin{figure*}[hbt]
\setlength{\tabcolsep}{1pt}
\centering
\footnotesize
\resizebox{1.0\linewidth}{!}
{
\centering
\begin{tabular}{cccccc}
\includegraphics[width=0.14\linewidth]{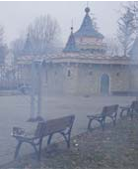}&
\includegraphics[width=0.14\linewidth]{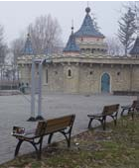}&
\includegraphics[width=0.14\linewidth]{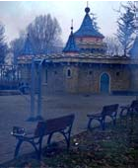}&
\includegraphics[width=0.14\linewidth]{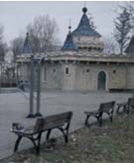}&
\includegraphics[width=0.14\linewidth]{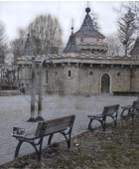}&
\includegraphics[width=0.14\linewidth]{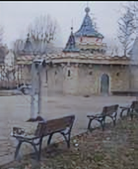}\\
 Input & GT & DCP~\cite{he2010single}& DCPDN~\cite{zhang2018densely} &GridDehazeNet~\cite{liuICCV2019GridDehazeNet} & \textbf{PRdehaze}(ours)\\
\end{tabular}
}
\caption{Dehazing comparisons for the O-HAZE images.}
  \label{fig:Ohazy}
\end{figure*}

\vspace{-0.5cm}
\subsection{Real Dataset}
O-HAZE Dataset~\cite{ancuti2018haze} consists of real haze-free and hazy images which include 45 different outdoor scenes. The images are generated by professional haze machines under the same illumination parameters, which gives a good reference for the dehazing performances.
The proposed model is trained using the official train set and evaluated it on the test set in the same way as the compared methods~\cite{dong2020multi}. Note there is no groundtruth transmission map, our model is only supervised by the haze-free images.

The proposed model obtains the best performances for the real hazy images which demonstrates the practicability for the real applications. Interactively optimizing the physical model parameter with progressive residual learning strategy improves the adaptability of the methods. The visual results of the real cases are compared in Fig.\ref{fig:Ohazy}.

\section{Discussions and conclusions}
\textbf{Progressive refinement of the dehazed image and residuals by the physical model-free components.}
The dehazed images are progressively restored by the physical model-free components. In our implementations, 3 iterations are applied and quantitative PSNR / SSIM comparisons on Cityscapes dataset for image dehazing estimations in each iteration are provided in Table \ref{tab:iter}.

It is clear that our progressive residual learning strategy has improved the dehazing performances step by step. With better dehazing estimations, more accurate residuals information would be obtained and fed into the module, which in turn would benefit the learning of the dehazing process. 

\textbf{Synthetic vs. real hazy data.}
 In Table \ref{tab:dehaze}, it is interesting that the initial physical model-free component achieves better objective assessment index than the following physical model-based process for synthetic dataset SOTS and Cityscapes, where the data is easy to obtain. However, the following physical model-based component achieves much better results than the initial physical model-free component for real hazy dataset such as O-HAZE dataset where the data is rather limited and valuable. It proves that with enough training data, physical model-free methods could outperform the physical model-based methods as the physical model-based methods perform the multi-objective optimizations of model parameters which may lead to sub-optimal learning of final dehazing task. However, physical model-free methods suffer from the limited training data problem in real applications, where physical model-based methods perform better with the guidance of the physical laws.  Luckily, with the residual learning, our method can enjoy the merits of methods in both categories and performs favorably against the state-of-the-art methods. Global atmospheric light and transmissions map are interactively optimized and closely related by the residuals learning, which brings better model interpretability and adaptivity for complex foggy data.
\vspace{-0.5cm}
\begin{table}[hbt]
\setlength{\tabcolsep}{2.5pt}
\begin{center}
 \caption{Quantitative PSNR / SSIM comparisons on O-HAZE dataset.}
  \label{tab:O-HAZE}
    \tiny
  \begin{tabular}{ccccccccc}
    \hline
                 &\textbf{DCP}~\cite{he2010single}&\textbf{AOD-Net}~\cite{li2017aod}& \textbf{MSCNN}~\cite{ren2016single}&\textbf{GCANet}~\cite{chen2018gated}&  \textbf{GFN}~\cite{ren2018gated}& \textbf{MSBDN}~\cite{dong2020multi} &  \textbf{PRdehaze}\\
    \hline
    PSNR / SSIM &16.78 / 0.653&15.03 /0.539&17.56/ 0.650&16.28 / 0.645&18.16 / 0.671&24.36 / 0.749&\textbf{24.61 / 0.771}\\
    \hline
  \end{tabular}
  \end{center}
\end{table}
\vspace{-1cm}
\begin{table}[htb]
\begin{center}
 \caption{Quantitative PSNR / SSIM comparisons on Cityscapes dataset for image dehazing estimations in each iteration by the physical model-free components.}
  \label{tab:iter}
    \scriptsize
  \begin{tabular}{cccc}
    \hline
              & iter1 & iter2 & iter3  \\
    \hline
    PSNR / SSIM & 23.57/0.928 & 26.11/0.957 & 26.64/0.965   \\
    \hline
  \end{tabular}
  \end{center}
\end{table}
\vspace{-1cm}
\begin{table}[htb]
\begin{center}
 \caption{Quantitative PSNR / SSIM comparisons on SOTS, Cityscapes and O-HAZE dataset for dehazing estimations by the physical model-free component only, as well as the following reformulated scattering model-based components with preliminary or refined transmissions map estimated.}
  \label{tab:dehaze}
    \scriptsize
  \begin{tabular}{cccc}
    \hline
              & model-free & preliminary combining  & refined  \\
    \hline
    SOTS    &33.06 / 0.9802  & 32.96 / 0.9798&  29.41 / 0.9692  \\
    Cityscapes &28.20 / 0.9713   & 27.75 / 0.9698&  26.77 / 0.9694  \\
    O-HAZE&24.16 / 0.7477   & 24.15 / 0.7575&  24.61 / 0.7706  \\
    \hline
  \end{tabular}
  \end{center}
\end{table}

%
\vspace{-0.5cm}
In this paper, a progressive residual learning strategy has been proposed to combine the physical model-free dehazing process with reformulated scattering model-based dehazing operations. Global atmosphere light and transmission maps are interactively optimized and closely related by the residuals information. The physical parameters learn better predictions with the accurate residual information and preliminary dehazed restorations from the initial physical model-free dehazing process. The proposed method compares favorably against the state-of-the-art methods on indoor, outdoor and real hazy image dataset with better model interpretability and adaptivity.

\bibliographystyle{IEEEbib}
\bibliography{icmehaze}

\end{document}